\documentclass[pre,tighten]{revtex4}
\usepackage[latin1]{inputenc}
\usepackage{amssymb,amsmath}
\usepackage{graphicx}

\begin{document}
\title{Order statistics of Rosenstock's trapping problem \\
in disordered media}
\author{S. B. Yuste}
\email[E-mail: ]{santos@unex.es}
\author{L. Acedo}
\email[E-mail: ]{acedo@unex.es}
\affiliation{%
Departamento de F\'{\i}sica, Universidad  de  Extremadura, E-06071
 Badajoz, Spain
}%

\date{\today}
\begin{abstract}
The distribution of times $t_{j,N}$ elapsed until the first $j$
independent random walkers from a set of $N \gg 1$, all starting
from the same site, are trapped by a quenched configuration of
traps randomly placed on a disordered lattice is investigated. In
doing so, the cumulants of the distribution of the territory
explored by $N$ independent random walkers $S_N(t)$ and the
probability $\Phi_N(t)$ that no particle of an initial set of $N$
is trapped by time $t$ are considered. Simulation results for the
two-dimensional incipient percolation aggregate show that the
ratio between the $n$th cumulant and the $n$th moment of $S_N(t)$
is, for large $N$, (i) very large in comparison with the same
ratio in Euclidean media, and (ii) almost constant. The first
property implies that, in contrast with Euclidean media,
approximations of order higher than the standard zeroth-order
Rosenstock approximation  are required to provide a reasonable
description of the trapping order statistics. Fortunately, the
second property (which has a geometric origin) can be exploited to
build these higher-order Rosenstock approximations. Simulation
results for the two-dimensional incipient percolation aggregate
confirm the predictions of our approach.
\end{abstract}
 \maketitle

\section{Introduction}
Rosenstock's trapping problem is a fundamental problem of random
walk theory that has been of interest for decades
\cite{TrappingN1,BluKlafZum,Hughes,Havlin}. Most studies refer to
the case in which a single ($N=1$) random walker  is placed
initially at a site of a Euclidean or disordered lattice which is
randomly filled with trap sites at a concentration $c$, and then
performs a random walk until it is absorbed by one of these traps.
The statistical quantity of main interest in this problem is the
survival probability $\Phi_1(t)$ that the random walker is not
trapped by time $t$, from which one can obtain the moments
 $\langle t^m \rangle=\int_0^\infty t^m [1-\Phi_1(t)]dt$ of the lifetime of this
random walker. This problem has its origin in Smoluchowski's
theory of coagulation of colloidal particles
\cite{Hughes,Havlin,Hollander} and has been applied to many
systems in physics and chemistry such as trapping of mobile
defects in crystals with point sinks
\cite{Beeler,Rosenstock,Damask}, the kinetics of luminescent
organic materials \cite{Rosenstock}, anchoring of polymers by
chemically active sites \cite{Oshanin}, and atomic diffusion in
glasslike materials \cite{Miyagawa}, among others.

A generalization of the trapping problem to the case of $N$
independent random walkers was studied by Krapivsky and Redner in
Ref.~\cite{KR}. In particular, they studied the problem of $N$
diffusing predators placed initially at a given distance from a
diffusing prey in one dimension. The model in which static preys
are stochastically distributed all to one side of the predators
was the subject of a later analysis \cite{OneSided}. Shortly
after, the order statistics of the trapping problem in
$d$-dimensional Euclidean lattices for a set of $N\gg 1$
independent random walkers, i.e., the statistical description of
the time $t_{j,N}$ elapsed until the first $j$ out of $N$
independent random walkers (initially starting at the same site)
are trapped by quenched traps randomly arranged on $d$-dimensional
Euclidean lattices, was studied (and rigorously solved for the
one-dimensional case) in Ref.~\cite{PREost}. In this work, the
moments $\langle t_{j,N}^m \rangle$, $m=1,2,3,\ldots$ were
calculated from the probability $\Phi_{j,N}(t)$ that $j$ random
walkers of the initial set of $N$ have been absorbed by time $t$.
The key step in this calculation was the assumption of
independency of the random walkers that allowed the establishment
of a relationship between $\Phi_{j,N}(t)$ and the survival
probability $\Phi_{N}(t)\equiv \Phi_{0,N}(t)$ of the full set of
$N$ random walkers \cite{PREost}. The survival probability
$\Phi_{N}(t)$ was calculated by means of Rosenstock's
approximation which required the evaluation of the first moment of
the number $S_N(t)$ of different sites visited (territory
explored)  by $N$ random walkers.

Interest in multiparticle diffusion problems has had a boost
lately because of some advances in optical spectroscopy
\cite{SingMol} that make it possible to monitor events
corresponding to single particles of an ensemble. The simultaneous
tracking of $N\gg1$ fluorescently labelled particles and the
analysis of the motions of the particles allows the study of local
conditions (mechanical response, visco-elasticity) inside many
complex structures such as fibrous polymers and the intracellular
medium \cite{MultiTracking}. But biological samples and many real
inorganic substances are disordered media (as opposed to
translationally invariant Euclidean media) which are usually
described as stochastic fractals
\cite{Havlin,FractalBooks,Pfeifer}. There are two main reasons for
this identification: disordered systems share the statistical
fractal structure of stochastic fractal models and diffusion is
anomalous in both media.

The single-particle ($N=1$) Rosenstock  trapping problem in
fractal media have been thoroughly discussed by Blumen, Klafter
and Zumofen \cite{BluKlafZum,BluKlafZumPRB}. In this paper we will
study its \emph{multiparticle } ($N\gg 1$) version, which is
relevant for all those cases where the diffusing particles are
placed (or created) in bunches. This may be especially important
if the first or first few particles that are absorbed lead to a
trigger effect. Here, we generalize to disordered fractal
substrates the results for the order statistics of the
multiparticle trapping problem obtained for Euclidean media in
Ref.~\cite{PREost}.  We will discover that for the two-dimensional
incipient percolation aggregate, and in sharp contrast with the
Euclidean media results, the zeroth-order Rosenstock approximation
is quite incapable of describing the survival probabilities, and
therefore the order statistics of the multiparticle trapping
problem we are dealing with.  This is because the ratio between
the cumulants $\kappa_m$ of the distribution of $S_N(t)$ and
$\langle S_N(t) \rangle^m$ is relatively large (and almost
constant) for $N\gg 1$. We traced the origin of this behavior to
the fact that the fluctuations in the number $S_N(t)$ of distinct
sites explored by a large number $N$ of random walkers are
negligible relative to the fluctuations in the number of sites
that form the stochastic substrate. As a practical consequence,
higher-order Rosenstock approximations are required for these
media in order to give an account of the order statistics trapping
problem with an accuracy similar to those reached by means of the
zeroth-order Rosenstock approximation for Euclidean lattices. The
idea of evaluating the survival probability for the multiparticle
trapping problem for Euclidean media by means of the Rosenstock
approximation was first suggested by Larralde \emph{et al.\ } in
\cite{LarraldePRA}, although, to the best of our knowledge, it has
not been implemented (except for the so-called ``one-sided
trapping problem'' \cite{OneSided}) perhaps for the lack of
precise expressions for the moments $\langle S_N^m(t) \rangle$ of
$S_N(t)$. However, for the percolation aggregate, we will discover
in Sec.\ \ref{sec:SNtmom} that  one can get a very good estimate
of $\langle S_N^m(t) \rangle$ from the value of the first moment
$\langle S_N(t) \rangle$. This is fortunate because the asymptotic
expansion of $\langle S_N(t) \rangle$ for large $N$ is known for
this medium \cite{SNtFrac}.

The multiparticle Rosenstock trapping problem we are considering
here can be seen as a  stochastic generalization of the problem of
the order statistics for the sequence of trapping times (or exit
times) of a set of $N$ independent random walkers, all starting
form the same site at the same time, when the traps form a
``spherical'' absorbing boundary with a fixed radius. This problem
was first studied by Lindenberg {\em et al.\ } \cite{Lindenberg}
and Weiss {\em et al.\ } \cite{WeisShulLind} for Euclidean
lattices (mainly for the one dimensional case). After these
pioneering works, improved results and extensions to deterministic
and random fractal substrates have been reported
\cite{trapesfera,DragerKlafter}.  A closely related  multiparticle
Brownian problem has been recently considered by Bénichou \emph{et
al.} \cite{BenichouJPA03}: they studied the join residence times
of $N$ independent Brownian particles in a disc of a given radius.
In particular, they studied the time spent by all $N$ particles
simultaneously in the disc within a given time interval, and the
time which at least $N-j$ out of $N$ particles spend together in
the disc within a time interval.

 The plan of the paper is as follows. In
Sec.\ \ref{sec:Ost} the expressions that describe the order
statistics of the trapping process are deduced. In Sec.\
\ref{sec:SNtmom} we study the moments $\langle S_N^m(t)\rangle$ of
the territory explored by $N$ independent random walkers  on
two-dimensional incipient percolation aggregates by means of
numerical simulation. The results of Sec.\ \ref{sec:SNtmom} are
applied in Sec.\ \ref{sec:Rosen} to obtain the survival
probability $\Phi_N(t)$ by means of Rosenstock's approximation.
Then we calculate the moments $\langle t_{j,N}^m \rangle$ of the
time elapsed until the first $j$ random walkers are trapped for
every $j=1,2,\ldots$ and $m=1,2,\ldots$ and compare these
predictions with simulation results for the two-dimensional
incipient percolation aggregate. A general discussion and
conclusions are given in Sec.\ \ref{sec:Conclu}.

\section{Definitions and fundamental relations}
\label{sec:Ost} The results and definitions of this section have
already been discussed in detail in the context of the trapping
problem in Euclidean media \cite{PREost}. However, we will briefly
summarize  those results that are basic and necessary in order to
follow the arguments in the rest of the paper.

Let us first define $\Psi_{j,N}(t)$ as the probability that $j$
random walkers of the initial set of $N$ have been absorbed by
time $t$ by a given configuration of traps arranged on a given
realization of the disordered substrate. The quantity of
statistical interest is the average $\Phi_{j,N}(t)= \langle
\Psi_{j,N}(t) \rangle$ performed over all the possible outcomes of
the ``trapping experiment'' carried out in a quenched
configuration of traps in a given lattice realization followed by
an average over all trap configurations and lattice realizations.
We will also denote by $\Psi(t)$ the probability that a single
random walker has not been absorbed by time $t$ in this quenched
configuration of traps placed upon a specific lattice realization.
This is commonly known as the survival probability. It is then
clear that
\begin{equation}
\Psi_{j,N}(t)=  \binom{N}{j}  \left(1-\Psi\right)^j  \Psi^{N-j}=
\binom{N}{j} \sum_{m=0}^j (-1)^m \binom{j}{m} \Psi^{N-j+m}=
 (-1)^j  \binom{N}{j} \nabla^j \Psi_{0,N}
\label{PsijN}
\end{equation}
where $\nabla^j \Psi_{0,N}(t)=\sum_{m=0}^j (-1)^m \binom{j}{m}
\Psi_{0,N-m}(t)$ is just the backward difference formula for the
$j$th derivative of $\Psi_{0,N}(t)$, $d^j\Psi_{0,N}(t)/dN^j$.
Averaging over different configurations, and taking into account
that $\Phi_{0,N}(t)\equiv\Phi_N(t)= \langle\Psi_{0,N}(t)\rangle$
and $\Phi_{j,N}(t)= \langle\Psi_{j,N}(t)\rangle$, we get
\begin{equation}
\Phi_{j,N}(t) = (-1)^j \binom{N}{j}
  \nabla^j \Phi_N(t)   .
\label{PhijN}
\end{equation}

Let us call $\mathcal{S}_j$ the state in which $j$ particles have
been absorbed and $N-j$ particles of the initial set of $N$ are
still diffusing, and let $h_{j,N}(t)\, dt$ be the probability that
the $j$th absorbed particle of the initial set of $N$ disappears
during the time interval $(t,t+dt]$. The change of the probability
of $\mathcal{S}_j$ during the time interval $(t,t+dt]$ is given by
$\Phi_{j,N}(t+dt)-\Phi_{j,N}(t)$. But this probability changes
during this time interval by two causes: first, because the state
$\mathcal{S}_{j-1}$ can become the state $\mathcal{S}_{j}$ if a
particle of the set of $N-j+1$ particles still diffusing is
trapped during the time interval $(t,t+dt]$ [which happens with
probability $h_{j,N}(t) dt$] and, second, because
$\mathcal{S}_{j}$ can become the state $\mathcal{S}_{j+1}$ if a
particle of the $N-j$ particles still diffusing is trapped during
this time interval  [which happens with probability $h_{j+1,N}(t)
dt$]. Therefore
$\Phi_{j,N}(t+dt)-\Phi_{j,N}(t)=[h_{j,N}(t)-h_{j+1,N}(t)] dt$,
i.e.,
\begin{equation}
h_{j+1,N}(t)=h_{j,N} -\frac{d}{dt}\, \Phi_{j,N}(t) \label{hj1N}
\end{equation}
with $h_{0,N}=0$.
Then, the $m$th moment of the time in which the $j$th particle is trapped is given by
\begin{equation}
\langle t_{j,N}^m  \rangle = \int_0^\infty t^m h_{j,N}(t) dt \;.
\label{tjN}
\end{equation}
Using Eqs.\ (\ref{PhijN}), (\ref{hj1N}) and (\ref{tjN}) we find
the \emph{exact} recursion relation
\begin{equation}
\langle t_{j+1,N}^m \rangle = \langle t_{j,N}^m \rangle + (-1)^j
\binom{N}{j}
 \nabla^j \langle t_{1,N}^m \rangle
\label{tjNb}
\end{equation}
where
\begin{equation}
\langle t_{1,N}^m \rangle = m \int_0^\infty t^{m-1} \Phi_{N}(t) dt \; .
\label{t1Na}
\end{equation}
The set of Eqs.\ (\ref{tjNb}) and (\ref{t1Na}) is remarkable
because it implies that the order statistic of the trapping
problem can be described from the knowledge of $\langle t_{1,N}^m
\rangle $ only.  The difference derivative  $\nabla^j $ in Eqs.\
\eqref{PhijN} and \eqref{tjNb}  can be approximated by the
ordinary derivative  $d^j /dN^j$ when $j \ll N$. This will be
justified (and used) in  Sec.\ \ref{sec:Rosen}.

\section{Moments of the territory explored by $N\gg 1$ random walkers on
a two-dimensional incipient percolation aggregate}
\label{sec:SNtmom}The diffusion in percolation clusters as a model
of transport in disordered media was first suggested by de Gennes
\cite{Gennes}. Percolation clusters are disordered fractals: they
share the self-similarity property with deterministic fractals
build up through deterministic rules but only in a statistical
sense. In order to characterize these fractals several static and
dynamic exponents have been defined. Perhaps, the most widely
known is the fractal dimension, $d_f$, which, in the case of
disordered systems, is more conveniently defined using the scaling
of mass with linear size, $M \sim L^{d_f}$. However, random
walkers in disordered structures are forced to follow the paths
formed by the bonds between sites and, consequently, it is more
natural to define a chemical (or topological) distance between two
sites as the length of the shortest path along lattice bonds,
$\ell$. If we consider the number of sites inside an hypersphere
of radius $\ell$, $V(\ell)$, usually known as chemical volume
(also coincides with the mass if we assume that every site has a
unit mass) it is expected that $\langle V(\ell) \rangle \sim V_0
\ell^{d_\ell}$, where the brackets refers to an average over all
possible realization of the lattices and $d_\ell$ is the chemical
dimension. Similarly, the generalized Einstein's law of diffusion
for anomalous systems can be written in terms of the ordinary
Euclidean distance, $r$, or the chemical one, $\ell$, and we have:
\begin{equation}
\begin{array}{rcl}
\langle r^2 \rangle &\sim&  2 D t^{2/d_w}\, ,\\
\noalign{\smallskip} \langle \ell^2 \rangle &\sim&  2 D_\ell
t^{2/d_w^{\ell}}\;  ,
\end{array}
\end{equation}
for $t\gg 1$, and where $D$ and $D_\ell$ are the diffusion
coefficient and the chemical diffusion coefficient, respectively.
The exponent $d_w$ is the random walk dimension, also known as
diffusion exponent. The exponent $d_w^\ell$ corresponding to the
chemical (or topological) metric is called chemical random walk
dimension. Another important exponent appearing in the theory of
random walks in disordered media is the spectral or fracton
dimension, $d_s=2 d_f/d_w=2 d_\ell/d_w^{\ell}$ \cite{Havlin}.

\begin{figure}
\includegraphics[width=0.7 \columnwidth]{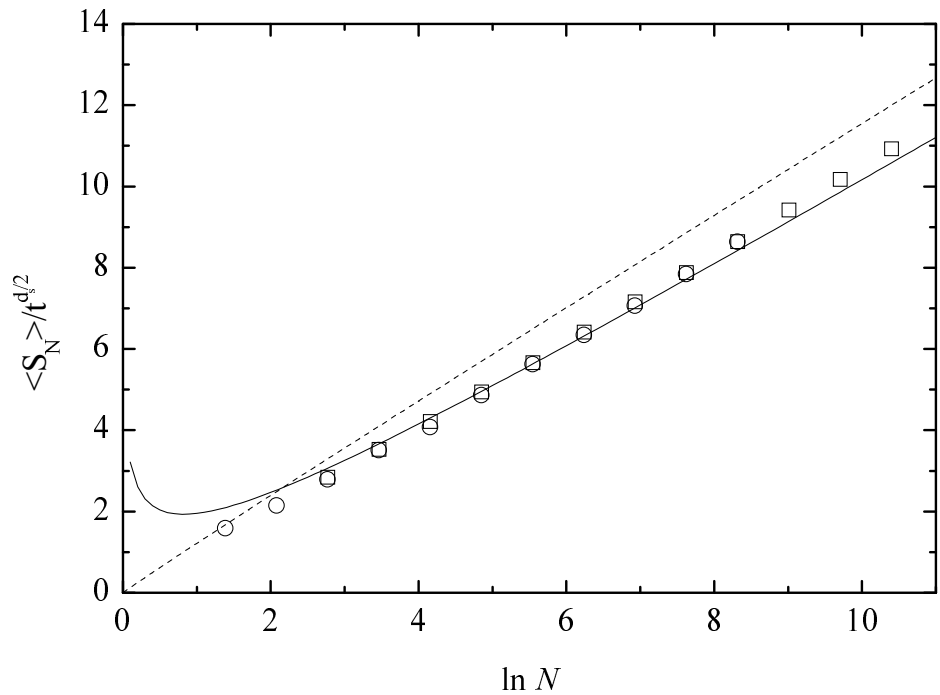}
\caption{$\langle S_N(t) \rangle/t^{d_s/2}$ versus $\ln N$ in the
two-dimensional incipient percolation aggregate. The lines
represent the result of the zeroth-order approximation (dashed
line) and the first-order approximation (solid line). The symbols
are simulation results obtained with 40000 experiments for
$t=1000$ (circles) and with 10000 experiments for $t=2000$
(squares). \label{fig1}}
\end{figure}

 Rosenstock's procedure \cite{Rosenstock} for
evaluating the survival probability of a set of random walkers
requires the knowledge of the first moments of the territory
explored $S_N(t)$ by these random walkers. This is an interesting
(and difficult) problem in itself that has already been thoroughly
studied in the case of $N\gg 1$ independent random walkers
\cite{KR,LarraldePRA,LarraldeNPO,DragerKlafter,SNtEuc,SNtFrac,MultiparticleReview}
although only the first moment $\langle S_N(t) \rangle $ has been
rigourously estimated \cite{SNtEuc,SNtFrac,MultiparticleReview}.
The average value of the territory explored by $N\gg 1$ random
walkers, all starting from the same site, in a disordered medium
was analyzed in Ref. \cite{DragerKlafter,SNtFrac} and it was found
that \cite{SNtFrac}
\begin{equation}
\langle S_N(t) \rangle \sim \bar{S}_N t^{d_s/2} \label{SNt1}
\end{equation}
with
\begin{equation}
\bar{S}_N = V_0 (2D_\ell)^{d_\ell/2} \left(\frac{\ln N}{\hat c}
\right)^{d_{\ell}/v} \left[1- d_\ell \frac{d_w^{\ell}-1
}{d_w^{\ell}} \displaystyle\sum_{n=1}^\infty\, \left(\ln
N\right)^{-n} \, \displaystyle\sum_{m=0}^n \, s_m^{(n)} \left(\ln
\ln N\right)^m \right] \, .\label{SNt}
\end{equation}
The parameters $\hat{c}$ and $s_m^{(n)}$ are characteristic of the
lattice and some of their values for several Euclidean and fractal
media are known \cite{SNtEuc,SNtFrac,MultiparticleReview}. In
particular, for the two-dimensional incipient percolation
aggregate, Monte Carlo simulations in this substrate (with
particles jumping from a site to one of its nearest neighbors
placed at one unit distance in each unit time)  have shown
\cite{SNtFrac,MultiparticleReview} that $d_{\ell} \simeq 1.65$,
$v=d_w^\ell/(d_w^\ell-1)$ with $d_w^\ell\simeq 2.45$,
$d_s=2d_\ell/d_w^\ell\simeq 1.35$, $V_0\simeq 1.1$, $\hat{c}
\simeq 1.05 $,  $2D_\ell \simeq 1.2$, $s_0^{(1)}=-\gamma-\ln \hat
A\hat c^{\hat\mu}\simeq  -0.62$ ($\gamma\simeq 0.577216$ is the
Euler constant) and $s_1^{(1)}=\hat{\mu} \simeq 0.8$ (see Table
\ref{table1}). Hence we have a reasonable estimate of the
asymptotic series for $\langle S_N(t) \rangle$ in Eq.\ (\ref{SNt})
up to first order ($n=1$), which is sufficient to account for
simulation results, as Fig.\ \ref{fig1} shows.

\begin{table}
\caption{ Parameters appearing in the asymptotic expression of
$\langle S_N(t)\rangle$ and the ratios $\ell_2$ and $\ell_3$ for
the two-dimensional incipient percolation aggregate.\label{table1}
}
\begin{ruledtabular}
\begin{tabular}{ccccccccc}
$d_\ell$ & $d_w^\ell$ & $\hat A$ & $\hat c$ & $\hat \mu$ & $V_0$ & $D_\ell$ & $\ell_2$& $\ell_3$ \\
\colrule
1.65 & 2.45  & 1.0 &  1.05 & 0.8 & 1.1 & 0.6 & 0.14 &  0.015 \\
\end{tabular}
\end{ruledtabular}
\end{table}

\begin{figure}
\includegraphics[width=0.7 \columnwidth]{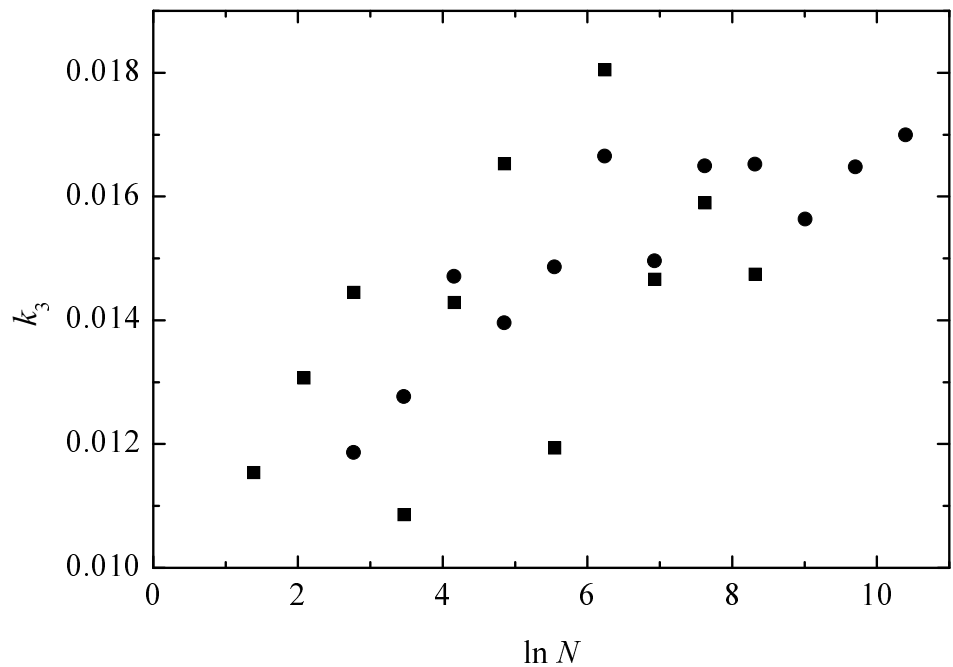}
\caption{Simulation values of $k_2=\kappa_2(S_N)/\langle S_N(t)
\rangle^2$  versus $\ln N$ for the two-dimensional incipient
percolation aggregate obtained averaging over $40000$ (squares)
and 20000 (circles) experiments for $t=1000$.\label{fig2}}
\end{figure}
\begin{figure}
\includegraphics[width=0.7 \columnwidth]{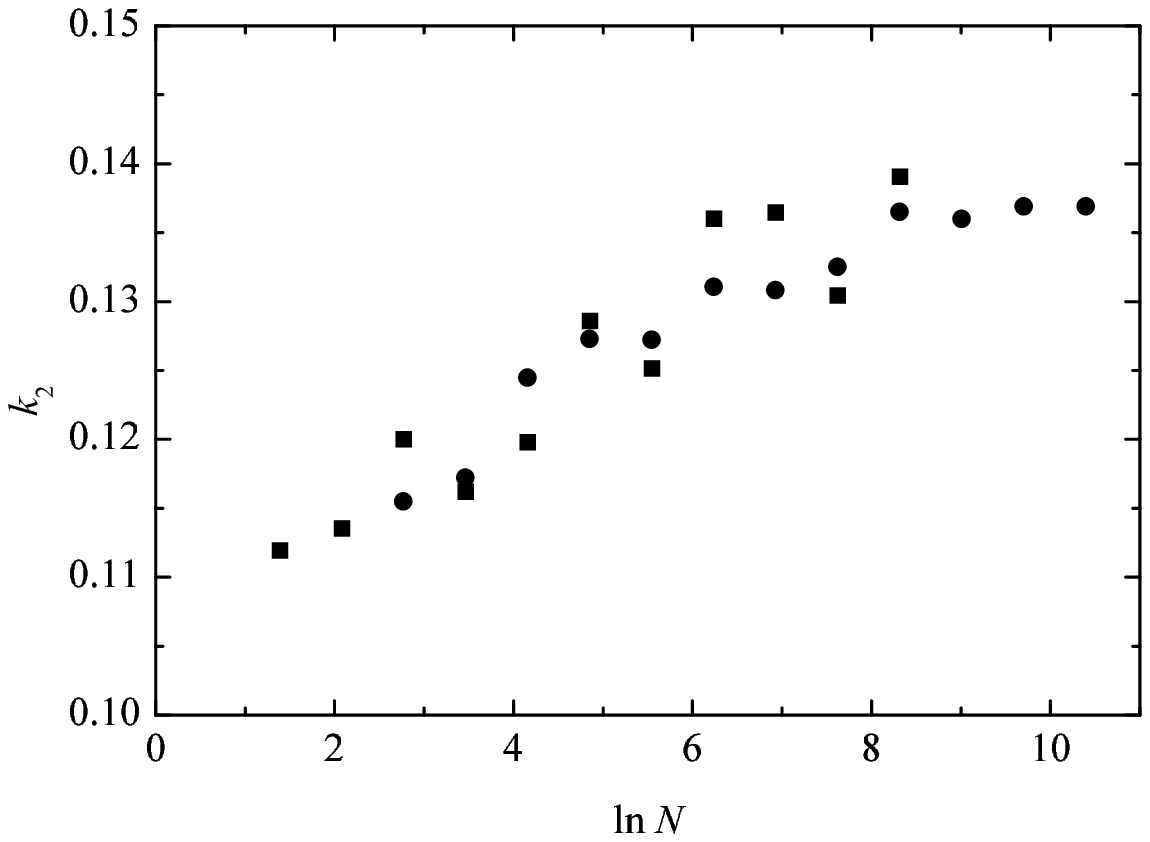}
 \caption{The
same as Fig.\ \ref{fig2} but for $k_3=\kappa_3(S_N)/\langle S_N(t)
\rangle^3$. \label{fig3}}
\end{figure}

\begin{figure}
\includegraphics[width=0.7 \columnwidth]{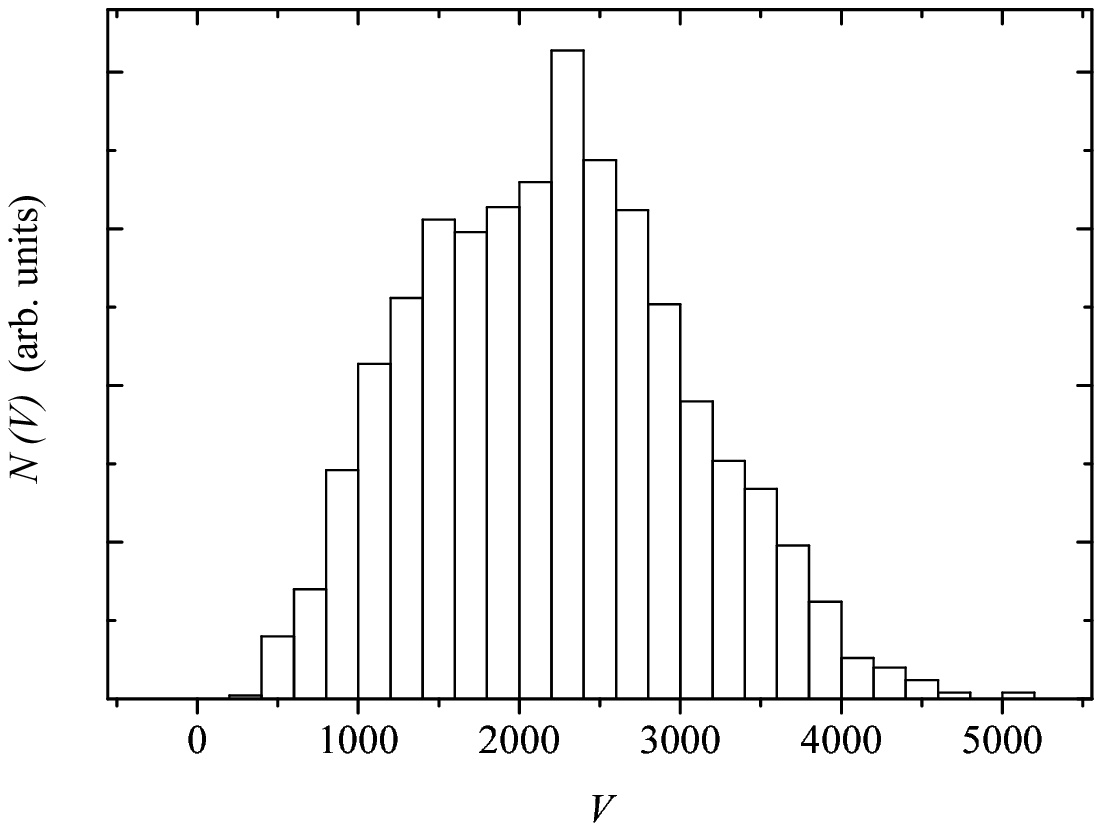}
\caption{Histogram for the chemical volume inside a circle of
chemical radius $\ell=100$ in the two-dimensional incipient
percolation aggregate. A set of $2000$ randomly generated clusters
in a $400\times 400$ square box were used in the computation. The
distribution is clearly asymmetric around the
maximum.\label{fig4}}
\end{figure}

In our simulations we also evaluated the second cumulant
(variance), $\kappa_2(S_N)=\langle S_N^2 \rangle-\langle S_N
\rangle^2$, and the third cumulant, $\kappa_3(S_N)=\langle S_N^3
\rangle-3\langle S_N^2\rangle \langle S_N \rangle+2 \langle S_N
\rangle^3$, of the territory explored $S_N(t)$ as they are
necessary for implementing the extended Rosenstock approximation
(see Sec.~\ref{sec:Rosen}). We found that the ratio
$k_m\equiv\kappa_m(S_N)/\langle S_N(t) \rangle^m$, although not
very sensitive to the value of $N$ (one can see in Figs.\
\ref{fig2} and  \ref{fig3} that these parameters are well
represented by $k_2=0.13\pm 0.02$ and $k_3=0.015\pm 0.02$ over a
wide range of $N$ values) seems to tend to a constant value for
large $N$ (about $0.14$ for $k_2$ and $0.016$ for $k_3$). This is
a surprising behavior that departs considerably from that of
 Euclidean media. For example, for the $d$-dimensional
Euclidean lattices it was found that $k_2$ goes as $1/\ln^2 N$ for
large $N$. Also, the value of $k_2$ for large $N$ is much smaller
than for the percolation aggregate (for example, for $N=2^{10}$,
$k_2 \simeq 1/15^2, 1/30^2$ and $1/50^2$ in the one-, two- and
three-dimensional Euclidean lattices, respectively \cite{PREost}),
which has important consequences for the accuracy of Rosenstock's
approximations of different orders, as we will show in
Sec.~\ref{sec:Rosen}. The disordered nature of the substrate must
be the reason for these remarkable differences in the behavior of
$k_m$. What is happening is that, for large $N$, the fluctuations
in the number $S_N(t)$ of distinct sites explored by a large
number $N$ of random walkers are dominated by the fluctuations
(over the set of stochastic lattice realizations used in the
simulations) of the number of sites inside a hypersphere of
chemical radius $\ell=\sqrt{2 D_{\ell}}\, t^{1/d_w^{\ell}}$. We
summarize this claim in a conjecture as follows
\begin{equation}
\lim_{N\rightarrow \infty} k_m=\lim_{N\rightarrow \infty}
\frac{\kappa_m(S_N(t))}{\langle S_N(t)
\rangle^m}=\frac{\kappa_m(V)}{\langle V(\ell) \rangle^m} \equiv
\ell_m \; , \label{conj}
\end{equation}
where $V(\ell)$ is the chemical volume (number of sites) of a
hypersphere of chemical radius $\ell$ and $\kappa_m(V)$ is the
$m$th-cumulant of the distribution of $V$. Rigorously, the
distance $\ell$ appearing in Eq.\ (\ref{conj}) is given by
$\sqrt{2 D_{\ell}}\, t^{1/d_w^{\ell}}$ which is the radius of the
diffusion front in the thermodynamic limit ($N \rightarrow
\infty$). However, the quotient $\ell_m$ is not very sensitive to
$\ell$ if a sufficiently large value of $\ell$ is taken. In Fig.\
\ref{fig4} we  plot a histogram for the chemical volume of a
two-dimensional incipient percolation aggregate with $\ell=100$,
evaluated using $2000$ realizations of the lattice. Thereby we
find that $\ell_2 \simeq 0.14$ and $\ell_3 \simeq 0.015$ in very
good agreement with the values of $k_2$ and $k_3$, respectively,
for large $N$ (see figures~\ref{fig2} and \ref{fig3}).
Consequently, we conclude that the fluctuations in $S_N(t)$ are
dominated by the disorder of the substrate and the influence of
the value of $N$ is completely overshadowed. Similar arguments
were presented by Rammal and Toulouse in their pioneer work on
percolation clusters \cite{RT}.

\section{Order statistics of the trapping process}
\label{sec:Rosen} Assume that one has a quenched configuration of
traps randomly placed on a given realization of the disordered
lattice with probability $c$. If $N$ random walkers start from an
origin site free from traps at $t=0$, it is clear that the
probability that all random walkers survive by time $t$ is given
by $(1-c)^{S_N(t)}$. The average of this quantity  over all
possible random walks, trap configurations, and substrate
realizations is known as the survival probability:
$\Phi_N(t)=\langle (1-c)^{S_N(t)} \rangle$. Using a well-known
theorem in statistics, we can define the $n$th-order Rosenstock
approximation $\Phi_N^{(n)}(t)$ for estimating $\Phi_N(t)$ as
\begin{equation}
\label{Ros:general} \Phi_N^{(n)}(t)=\exp\left[\displaystyle
\sum_{j=1}^{n+1} \frac{(-\lambda)^j}{j!} \kappa_j(S_N)  \right]
\end{equation}
where $\lambda \equiv -\ln (1-c)$ and $\kappa_j(S_N)$ is the
$j$th-cumulant of the distribution of the territory explored. In
the limit $n \rightarrow \infty$ we recover the exact result for
$\Phi_N(t)$. In the case of the single particle ($N=1$) trapping
problem, Eq.\ (\ref{Ros:general}) is known as the extended
Rosenstock approximation or truncated cumulant expansion
\cite{Hughes,Havlin,Hollander,BluKlafZumPRB,ZumofenBlumen,Weiss}.
Its generalization to the $N$ particle case was used in a
one-dimensional trapping problem in Ref.~\cite{OneSided}.

\begin{figure}
\includegraphics[width=0.7 \columnwidth]{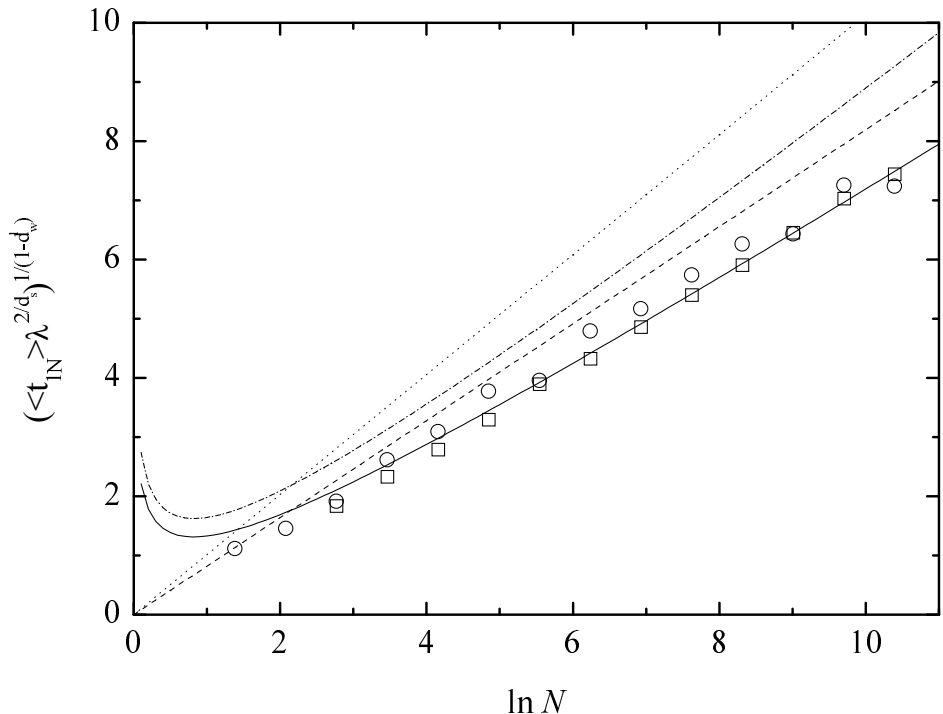}
\caption{$\left(\langle t_{1,N} \rangle \lambda^{2/d_s}
\right)^{1/(1-d_w^\ell)}$ versus $\ln N$ for the two-dimensional
incipient percolation aggregate. The lines represent the
$n$th-order Rosenstock approximation that uses the $l$th-order
approximation for $S_N(t)$ with, from top to bottom, $n=0$ and
$l=0$ (dotted line), $n=0$ and $l=1$  (dashed-dotted line), $n=2$
and $l=0$ (dashed line), and $n=2$ and $l=1$  (solid line). In
this and the following figures  we have used $k_2=0.13$ and
$k_3=0.015$. The symbols represent simulation results for
$c=0.008$ (average over $2000$ lattice realizations; circles) and
$c=0.001$ ($20000$ lattice realizations; squares).\label{fig5}}
\end{figure}

In the previous section we showed that Monte Carlo simulations
strongly suggest that $\kappa_n(S_N) \simeq k_n \langle S_N(t)
\rangle^n$ for large $N$, where $k_n$ are constants. [Notice that
$\kappa_1(S_N)=\langle S_N(t) \rangle$, so that $k_1=1$.]
Therefore, inserting this result into Eq.\ (\ref{Ros:general}),
the $n$th-order Rosenstock approximation becomes
\begin{equation}
\label{Ros:general1} \Phi_N^{(n)}(t)\simeq \exp\left[\displaystyle
\sum_{j=1}^{n+1} \frac{(-\lambda)^j}{j!} \, k_j \, \langle S_N
\rangle^j \right]\;
\end{equation}
or equivalently, by using Eq.\ \eqref{SNt1},
\begin{equation}
\label{Ros:general2} \Phi_N^{(n)}(t)\simeq \exp\left[\displaystyle
\sum_{j=1}^{n+1} \frac{(-\lambda)^j}{j!} \,k_j \,\bar{S}_N^j\,
t^{j d_s/2}  \right]\; .
\end{equation}
We can now evaluate an approximation for the moments of the first
trapping time, $\langle t_{1,N}^m \rangle$, by means of Eq.\
(\ref{t1Na}) assuming  that the contribution of $\Phi_N(t)$ to
$\langle t_{1,N}^m \rangle$ is negligible for those times for
which  $\Phi_N(t)$ and  $\Phi_N^{(n)}(t)$ differ substantially.
Therefore, the substitution of Eq.\ (\ref{Ros:general2}) into Eq.\
(\ref{t1Na}) yields
\begin{equation}
\langle t_{1,N}^m \rangle \simeq
 \langle t_{1,N}^m \rangle_n \equiv
  m \int_0^\infty t^{m-1} \exp\left[\displaystyle
\sum_{j=1}^{n+1} \frac{(-\lambda)^j}{j!} k_j \bar S_N^j t^{j
d_s/2} \right] dt \; . \label{t1Nb1}
\end{equation}
Writing $v=\lambda \bar{S}_N t^{d_s/2}$,  the $n$th-order
Rosenstock's estimate $\langle t_{1,N}^m \rangle_n$ for $\langle
t_{1,N}^m \rangle$ becomes:
\begin{equation}
\langle t_{1,N}^m \rangle_n = \frac{2m}{d_s}\left(\lambda
\bar{S}_N\right)^{-2m/d_s} \tau_n(m) \label{t1Nb2}
\end{equation}
where
\begin{equation}
\tau_n(m) = \int_0^\infty v^{2m/d_s-1}  \exp\left[\displaystyle \sum_{j=1}^{n+1} \frac{(-1)^j }{j!} k_j \, v^j \right] dv \; .
\label{taunm}
\end{equation}
Therefore, we find that the different $n$th-order Rosentock
approximations $\langle t_{1,N}^m \rangle_n$ differ from each
 other only by a numerical factor $\tau_n(m)$ (an integral) that
depends only on the substrate through its spectral dimension $d_s$
and the set of parameters $k_j$, $j=1,2,\ldots$ that come from the
distribution of the chemical volume of this substrate [c.f. Eq.
\eqref{conj}]. The integral in Eq.\ (\ref{taunm}) is trivial for
$n=0$ and yields $\tau_0(m)=\Gamma(2 m/d_s)$. Using the values of
Table \ref{table1} we get $d_s=2 d_\ell/d_w^{\ell} \simeq 1.375$
and the estimates $\tau_0(1) \simeq 0.89$, $\tau_0(2) \simeq 1.95$
and $\tau_0(3) \simeq 10.9$ for the two-dimensional incipient
percolation aggregate. The integral in Eq.\ (\ref{taunm}) only
converges for even values of $n$ so the next meaningful
approximation corresponds to $n=2$. Taking the values $k_2=0.13$
and $k_3=0.015$ (which describe  $k_2$ and $k_3$ well over the
range of values of $N$ used in our simulations: see Figs.\
\ref{fig2} and \ref{fig3}) and evaluating the integral in Eq.\
(\ref{taunm}) numerically,  we found the second-order prefactors
$\tau_2(m)$: $\tau_2(1) \approx 1.24$, $\tau_2(2) \approx 6.04$
and $\tau_2(3) \approx 95.6$, which are systematically much larger
than the zeroth-order ones $\tau_0(m) $, especially when the order
$m$ of the moment is large. This means that Rosenstock's
approximations of order higher than zero must be necessary  to
provide reasonable predictions for $\Phi_N(t)$ and $\langle
t_{1,N}^m \rangle$ in disordered media, especially when the moment
$m$ is large.  It should be noticed that the expression for the
first trapping time in Eq.\ (\ref{t1Nb2}) includes two
approximations of different nature: (a)  one due to the fact that
we are using a finite number $n$ of terms in the cumulant
expansion [which only affects the factor $\tau_n(m)$]; and (b) the
other due to the finite number of terms considered for estimating
$\bar{S}_N$ by means of the asymptotic series \eqref{SNt}.
Consequently, it is convenient to classify these approximations by
indexing them with a pair of integers $(n,l)$: the first index
gives the order of the Rosenstock approximation that is used, and
the second gives the number of terms considered in the evaluation
of $\langle S_N(t) \rangle$. In this way, the approximation
$(n,0)$ corresponds to the replacement in Eq.\ \eqref{t1Nb2} of
$\bar{S}_N$ by the leading term of the series of Eq.\ (\ref{SNt}),
so that $\langle t_{1,N}^m \rangle_{n}=\langle t_{1,N}^m
\rangle_{n0}[1+O(1/\ln N)]$ with
\begin{eqnarray}
\langle t_{1,N}^m \rangle_{n0} &=& \frac{2m  \tau_n(m)}{d_s
\left[\lambda V_0 (2D_\ell)^{d_\ell/2}\right]^{2m/d_s} }
  \left( \frac{\ln N}{\hat c}\right)^{m(1-d_w^\ell)} \\
\noalign{\smallskip} & = &\frac{T_n(m)}{ \lambda^{2m/d_s} (\ln
N)^{m(d_w^\ell-1)}} \; , \label{t1Nmn0}
\end{eqnarray}
and where we have absorbed all the dependence on the lattice
characteristic parameters ($V_0$, $D_\ell$, $d_s$, \ldots) into
the coefficient $T_n(m)$. In the same way, if we take the two
first terms in the asymptotic series of  of Eq.\ (\ref{SNt}), we
find $\langle t_{1,N}^m \rangle_{n}=\langle t_{1,N}^m
\rangle_{n1}[1+O(1/\ln N)^2]$  where the approximation $(n,1)$ is
\begin{equation}
\langle t_{1,N}^m \rangle_{n1} =
 \frac{T_n(m)}{ \lambda^{2m/d_s}
(\ln N)^{m(d_w^\ell-1)}}
 \left[1- d_\ell \frac{d_w^\ell-1}{d_w^\ell} \frac{s_0^{(1)} +s_1^{(1)} \ln N}{\ln N}\right]^{-2m/d_s}
 \label{t1Nmn1}
\end{equation}
or, for $\ln N\gg 1$,
\begin{equation}
\langle t_{1,N}^m \rangle_{n1} =
 \frac{T_n(m)}{ \lambda^{2m/d_s}
(\ln N)^{m(d_w^\ell-1)}}
 \left[1+ m (d_w^\ell-1) \frac{s_0^{(1)} +s_1^{(1)} \ln N}{\ln N}\right]\;
 .
 \label{t1Nmn1b}
\end{equation}
In Fig.\ (\ref{fig5}) we compare simulation results for the
trapping time of the first particle with the theoretical
predictions given by Eqs.\ \eqref{t1Nmn0} and \eqref{t1Nmn1} when
the parameters of Table \ref{table1} are used. We see that the
second-order Rosenstock approximation leads to much better results
than the standard zeroth-order approximation.

\begin{figure}
\includegraphics[width=0.7 \columnwidth]{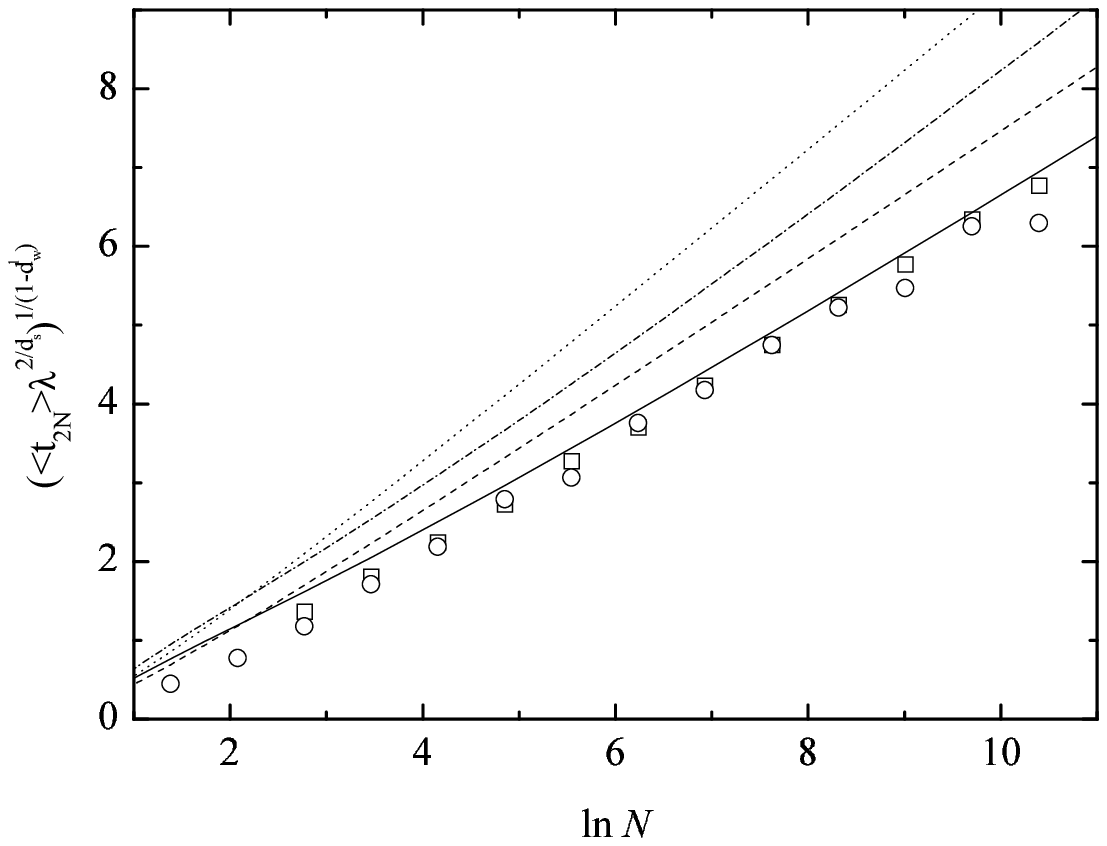} \caption{The
same as Fig.\ \ref{fig5} but for $\left(\langle t_{2,N} \rangle
\lambda^{2/d_s} \right)^{1/(1-d_w^\ell)}$.\label{fig6}}
\end{figure}

\begin{figure}
\includegraphics[width=0.7\columnwidth]{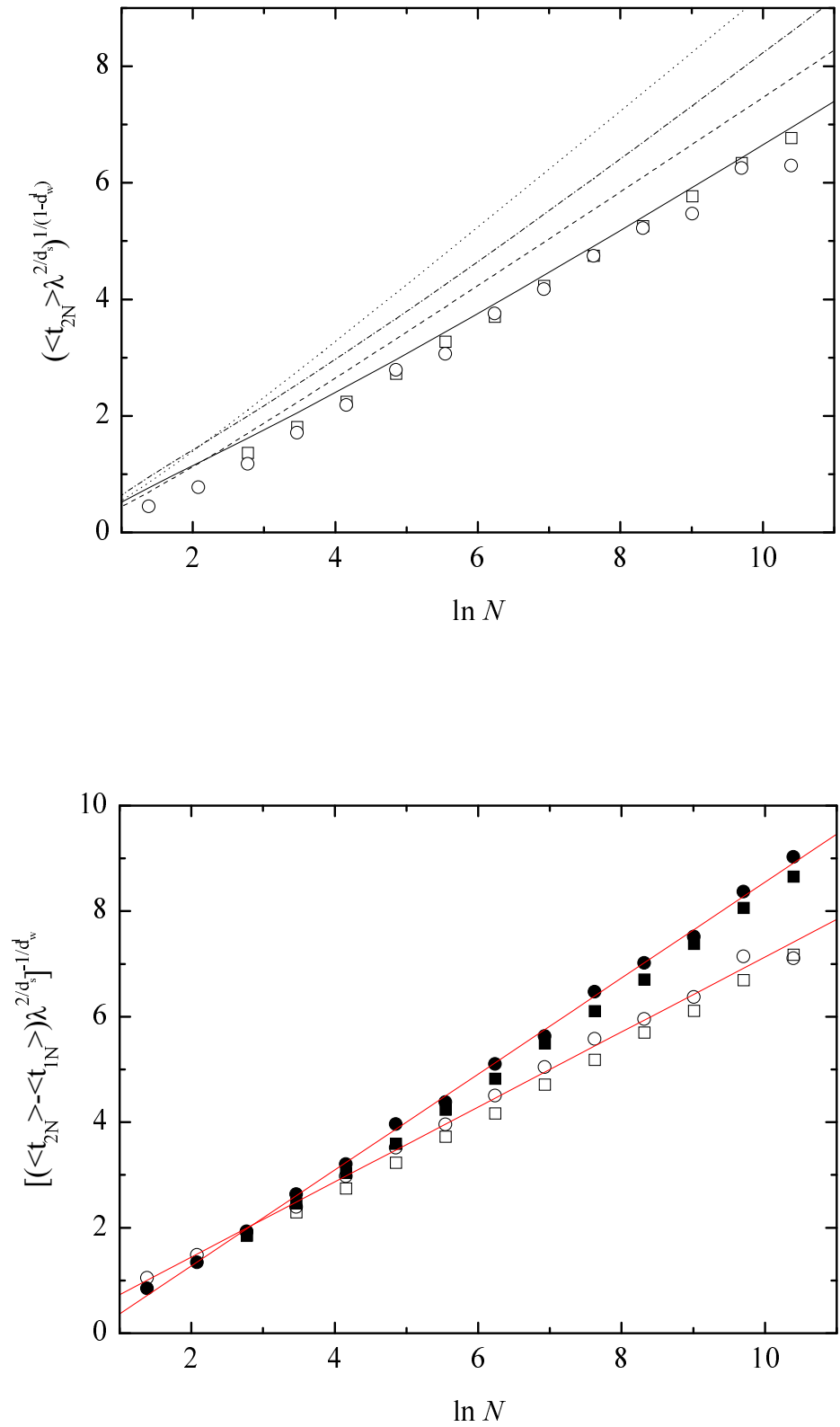} \caption{
Simulation results of $\left[\left(\langle t_{j+1,N}
\rangle-\langle t_{j,N} \rangle\right)\lambda^{2/d_s}
\right]^{-1/d_w^\ell}$ when $j=1$ (hollow symbols) and $j=2$
(filled symbols) for the two-dimensional incipient percolation
aggregate. Squares [circles] represent simulation results for
$c=0.001$ [$c=0.008$] averaged over $20000$ [$2000$] lattice
realizations. The lines are linear fits with slopes 0.72 (bottom)
and 0.90 (top).
 \label{fig7}}
\end{figure}

 The moments $\langle t_{j,N}^m \rangle$, $j=2,3,\ldots$
corresponding to the trapping of the $j$th particle absorbed by
the traps can be easily estimated by means of Eq.\ \eqref{tjNb}.
However, we can also obtain an explicit expression for $\langle
t_{j,N}^m \rangle$ if we approximate the difference operator
$\nabla^j$ in \eqref{tjNb} by the derivative  $d^j/dN^j$.   The
error in this approximation can be estimated from the equation
\begin{equation}\label{}
\frac{  \nabla^j f(N)}{(\Delta N)^j}=
  \frac{d^j f(N)}{dN^j}  + O\left(\frac{d^{j+1}f(N)}{
  dN^{j+1} } \, \Delta N
  \right).
\end{equation}
In our case $f(N)$ is $\langle t_{1,N}^m \rangle$ and $\Delta N=1$
so that
\begin{equation}\label{}
  \nabla^j \langle t_{1,N}^m \rangle=
  \frac{d^j\langle t_{1,N}^m \rangle}{dN^j} + O\left(\frac{d^{j+1}\langle t_{1,N}^m\rangle}{
  dN^{j+1}}
  \right).
\end{equation}
As
\begin{equation}
\label{djNjlnN} \frac{d^j}{dN^j} (\ln N)^{-\mu}=(-1)^j \mu
\frac{(j-1)!}{N^j}(\ln N)^{-\mu-1}+\frac{\ln \ln N}{N^j} {O}[(\ln
N)^{-\mu-2}]
\end{equation}
one finds, from Eq.\ (\ref{t1Nmn0}) [or Eq.\eqref{t1Nmn1}], that
\begin{equation}\label{}
\nabla^j \langle t_{1,N}^m \rangle=
  \frac{d^j\langle t_{1,N}^m \rangle}{dN^j}
  \left[1+O(N^{-1})\right].
\end{equation}
Taking into account that $\binom{N}{j}/N^j\simeq 1/j!$ for $j\ll
N$, we obtain from Eqs.\  (\ref{tjNb}), (\ref{t1Nmn0}) [or
\eqref{t1Nmn1}] and \eqref{djNjlnN} the recursion relation
\begin{equation}
\langle t_{j+1,N}^m \rangle = \langle t_{j,N}^m
\rangle+\frac{1}{j} m (d_w^\ell-1) T_n(m) \lambda^{-2m/d_s} (\ln
N)^{-m(d_w^\ell-1)-1} \left[1+O\left(\frac{1}{\ln N}
\right)\right]\;  \label{tjNc1}
\end{equation}
which can be easily solved:
\begin{equation}
\langle t_{j,N}^m \rangle=\langle t_{1,N}^m \rangle+ m(d_w^\ell-1)
T_n(m) \lambda^{-2m/d_s} \frac{\psi(j)+\gamma}{(\ln
N)^{m(d_w^\ell-1)+1}} \left[1+O\left(\frac{1}{\ln N}
\right)\right]\;  \label{tjNcPsi}
\end{equation}
where
\begin{equation}
\psi(j)=\psi(1)+\sum_{r=1}^{j-1} \frac{1}{r}
\end{equation}
is the psi (digamma) function \cite{abramo}, $\psi(1)=-\gamma$,
and $\gamma$ is the Euler constant. Equation \eqref{tjNcPsi}
yields
\begin{equation}
\langle t_{j,N}^m \rangle_{n0}=\frac{T_n(m)}{ \lambda^{2m/d_s}
(\ln N)^{m(d_w^\ell-1)}}
\label{tjNcPsib0}
\end{equation}
by using Eq.\ \eqref{t1Nmn0}, and
\begin{equation}
\langle t_{j,N}^m \rangle_{n1}=\frac{T_n(m)}{ \lambda^{2m/d_s}
(\ln N)^{m(d_w^\ell-1)}} \left[1+ m (d_w^\ell-1)
\frac{\psi(j)+\gamma+s_0^{(1)}  +s_1^{(1)} \ln N}
 {\ln N} \right]
\label{tjNcPsib1}
\end{equation}
when  Eq.\ \eqref{t1Nmn1b} is used. In Fig.\ (\ref{fig6}) we
compare the predictions for $\langle t_{2,N} \rangle$ obtained
from Eqs.\ (\ref{tjNcPsib0}) and (\ref{tjNcPsib1}) with simulation
results. The results are similar to that found in Fig.\
(\ref{fig5}) for $\langle t_{1,N} \rangle$. In Fig.\ (\ref{fig7})
the differences $\langle t_{j+1,N} \rangle - \langle t_{j,N}
\rangle$ estimated from Eq.\ (\ref{tjNc1}) are also plotted in a
scaled form for $j=1,2$. The theoretical prediction is that, for
large $N$, these points should tend to lie along a straight line
(which is true) with a slope $(d_w^\ell-1)T_2(1)/j$, i.e., a slope
$0.76$ for $j=1$ and 1.0 for $j=2$. The last prediction is not
good for $j=2$ (the fitted value is 0.90), but this should not be
surprising because in Eq.\ (\ref{tjNc1}) we have ignored
correction terms of order $1/\ln N$, which are very large even for
huge values of $N$. The only way to remedy this deficiency would
be by increasing the number of asymptotic terms retained in the
evaluation of $\langle S_N(t) \rangle$, which in turns requires
knowing $s_m^{(n)}$ for $n\ge 2 $ [c.f. Eq.\ \eqref{SNt}].
Unfortunately, these values of $s_m^{(n)}$ are very difficult to
estimate by means of numerical simulations \cite{SNtFrac} and are
unknown for $n\ge 2$.

\begin{figure}
\includegraphics[width=0.7 \columnwidth]{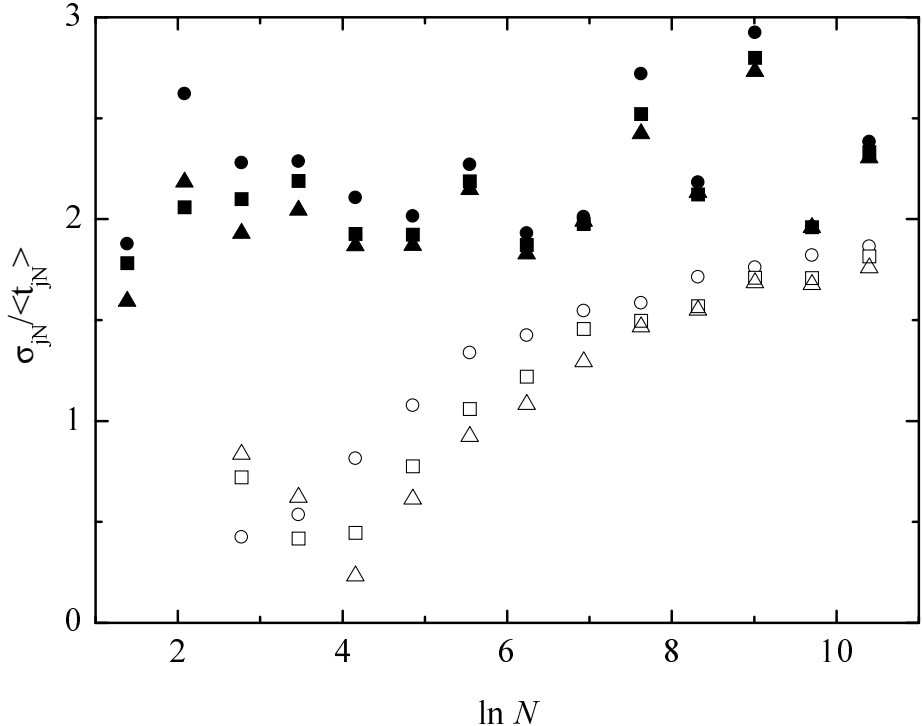} \caption{Simulation values of
$\sigma_{j,N}/\langle t_{j,N} \rangle$ versus $\ln N$ for $j=1$
(circles), $j=2$ (squares) and $j=3$ (triangles) and two trap
concentrations: $c=0.001$ (hollow symbols; $20000$ lattice
realizations) and $c=0.008$ (filled symbols; $2000$ lattice
realizations) for the two-dimensional incipient percolation
aggregate. The second order Rosenstock approximation predicts a
ratio close to $2$ for large values of $N$.\label{fig8}}
\end{figure}

The variance of $t_{j,N}$ is easily obtained from Eq.\
(\ref{t1Nb2}):
\begin{equation}
\sigma_{j,N}^2 = \frac{[T_n(2)-T_n^2(1)]}{ \lambda^{4/d_s} (\ln
N)^{2(d_w^\ell-1)}} \left[1+ 2 (d_w^\ell-1)
\frac{\psi(j)+\gamma+s_0^{(1)}  +s_1^{(1)} \ln N}
 {\ln N}+O\left(\frac{1}{\ln^2 N} \right)\right]\;
\label{sigmajN}
\end{equation}
and, consequently,
\begin{equation}
\frac{\sigma_{j,N}^2 }{\langle t_{1,N}\rangle^2 }= d_s
\frac{\tau_n(2)}{\tau_n(1)^2}-1\; . \label{ratiost2}
\end{equation}
This is an interesting result because it means that the ratio
between the variance of the first trapping time and the mean of
that time is, for large $N$, independent of $N$. The numerical
value of the ratio $\sigma_{j,N}/\langle t_{1,N} \rangle$ is $[d_s
\Gamma(4/d_s)/\Gamma^2(2/d_s)]-1\simeq 1.53$ for the zeroth-order
Rosenstock approximation, and $2.0$ for the second-order
approximation.  In Fig.\ \ref{fig8} we plot this ratio for
$j=1,2,3$ versus $\ln N$, and the second-order theoretical limit
$\sigma_{j,N}/\langle t_{1,N} \rangle \simeq 2$ seems to be
consistent with the simulation data.

Some considerations about the range of validity of the
approximations developed in this paper are called for at this
point. The approximation for $\langle S_N(t) \rangle$ in Eqs.\
(\ref{SNt1}) and (\ref{SNt}) is only valid in the so called Regime
II or intermediate time regime \cite{SNtFrac,MultiparticleReview}.
As the integral in Eq.\ (\ref{t1Na}) was evaluated assuming that
the expression for $\langle S_N(t) \rangle$ was valid for all
times we conclude that the integral of $m t^{m-1} \Phi_N(t)$ over
the short-time interval $[0,t_\times]$ ($t_\times \sim \ln N$
being the crossover time between Regime I and Regime II) has to be
negligible relative to $\langle t_{1,N}^m \rangle$, or
equivalently $(\ln N)^m \ll \langle t_{1,N}^m \rangle$, for our
approach and our results being reasonable. Taking into account the
estimate for $\langle t_{1,N}^m \rangle$ given in Eq.\
(\ref{t1Nmn0}), this condition can be written as $\lambda \ll (\ln
N)^{-d_\ell}$. The concentrations of traps we have used in our
simulations, $c=0.001$ and $c=0.008$, verify this condition for
all the values of $N$ considered. Apart from this upper bound on
$c$, we must also point out that, as also for Euclidean lattices,
our results break down if most of the trapping takes place within
the long-time Donsker-Varadhan regime \cite{DV}. Further reference
to this limitation of the theory presented in this paper will be
made below.

\section{Summary and conclusions}
\label{sec:Conclu} We have dealt with the following order
statistics problem: when $N$ independent random walkers all
starting from the same site diffuse on a disordered lattice
populated with a concentration $c$ of static trapping sites, what
is the distribution of the elapsed times, $t_{j,N}$, until the
first $j$ random walkers are trapped? We were able to generalize
the theory developed for the special case of Euclidean lattices
\cite{PREost} to the case of disordered substrates, and asymptotic
expressions for the moments $\langle t_{j,N}^m \rangle$ with $j\ll
N$ were obtained. To this end, we used the so-called Rosenstock
approximation, which is suitable for not very large times and
small concentrations of traps, $c$. In this approximation the
survival probability of the full set of $N$ random walkers,
$\Phi_N(t)$, is expressed in terms of the cumulants $\kappa(S_N)$
of the distribution of the territory covered $S_N(t)$.

Monte Carlo simulation results for $\kappa(S_N)$ in the
two-dimensional incipient percolation aggregate showed that for
large $N$ the ratio $k_m=\kappa_m(S_N)/\langle S_N(t) \rangle^m$
with $m=2,3$ hardly depends on $N$ and is very large in comparison
with the corresponding Euclidean ratio. We attribute this behavior
to the fluctuations in $S_N(t)$ being dominated by the
fluctuations in the volume of the medium inside a hypersphere of
chemical radius $\ell\sim \sqrt{2 D_\ell} \, t^{1/d_w^\ell}$. This
claim is supported by the result $k_m \approx l_m$ for $N\gg 1$
found by simulations of $S_N(t)$ and $V(\ell)$ in two-dimensional
incipient percolation aggregates, where $l_m=\kappa_m(V)/\langle
V(\ell)\rangle^m$ characterizes the fluctuations in the volume
$V(\ell)$. Therefore, the result $k_m \approx l_m$ for $N\gg 1$
implies that the fluctuations in $S_N(t)$ are mainly accounted for
by the fluctuations in $V(\ell)$, and that the fluctuations in
$S_N(t)$ induced by the randomness of the diffusion process are
irrelevant. One expects this also to be true for other disordered
media. Hence, if $\langle S_n(t) \rangle$ is known, the cumulants
of the distribution of $S_N(t)$ can be calculated (for any
sufficiently large value of $N$) from the cumulants of the
distribution of the chemical volume. Finally, taking into account
that $\langle S_N(t) \rangle$ is reasonably well known
\cite{SNtFrac}, we arrive at a closed expression for the survival
probability $\Phi_N(t)$ [c.f.\ Eq.\ \eqref{Ros:general1}] using
Rosenstock's approximation.  But from $\Phi_N(t)$ one gets the
probability $\Phi_{jN}(t)$ that $j$ random walkers of the initial
set of $N$ have been absorbed by time $t$ [c.f. Eq.\
\eqref{PhijN}], so that, finally, we get the moments of the
trapping times, $\langle t_{j,N}^m \rangle$ from the first moment
$\langle S_n(t) \rangle$ of the territory explored!

Comparison with simulation data shows that, in contrast with the
Euclidean case, Rosenstock's approximations of order higher than
zero are necessary to account for the order statistics results in
the two-dimensional percolation aggregate. This is a consequence
of the large value (in comparison with the Euclidean case) of
$k_m$, due to the large fluctuations in the territory explored by
the random walkers, induced, as we showed,  by the spatial
disorder of the substrate. However, some features of the order
statistics of trapping hold in the disordered case: for example,
we found that the ratio $\sigma_{j,N}/\langle t_{j,N} \rangle$
depends only on the lattice characteristic parameters $d_s$ and
$k_m$, $m=1,2,\ldots$ for large $N$. This was confirmed by
simulations in two-dimensional percolation aggregates.

There are some interesting problems that we still cannot answer
with the theory developed in this paper. For example, an important
quantity is the time $t_{N,N}$ elapsed until all the particles of
the inital set of $N$ are trapped. The evaluation of the moments
of this quantity would require specific techniques for $j \approx
N$ as our results are limited to the opposite limit $j \ll N$.
Moreover, the trapping of the last particles surely takes place in
the Donsker-Varadhan time regime \cite{DV} where Rosenstock's
approximation for the survival probability cannot be used. The
recent development of a Monte Carlo method to evaluate confidently
the survival probability in the Donsker-Varadhan time regime for
Euclidean lattices by Barkema {\em et al.\ } \cite{Barkema}
following a previous work of Gallos {\em et al.\ } \cite{Gallos}
could serve as starting point for tackling this problem. However,
one should be aware that this task is not a straightforward
generalization to disordered media of that carried out for
Euclidean lattices  because one has to take into consideration
that, as Shapir \cite{Shapir} pointed out, the Donsker-Varandhan
long-time behavior is dominated by the subset of lattice
realizations that are more ramified (with $d_s=1$). Consequently,
an efficient Monte Carlo technique to explore the relevant rare
lattice realizations in percolation clusters has to be devised
before.

\acknowledgments This work was supported by the Ministerio de
Ciencia y Tecnolog\'{\i}a (Spain) through Grant No. BFM2001-0718.


\end{document}